\documentclass[12pt,english]{paper}
\usepackage{amssymb}
\usepackage[T1]{fontenc}
\usepackage[latin1]{inputenc}
\usepackage{graphicx}

\makeatletter

\makeatletter

\makeatletter

\usepackage{geometry}

\makeatletter

\usepackage{epsfig}

\makeatother

\makeatother

\makeatother

\usepackage{babel}
\makeatother
\begin{document}

\title{Evaluation of the neutron background in cryogenic Germanium target for WIMP direct detection when using a reactor neutrino detector as a neutron veto system}

\author{Ye Xu%
\thanks{Corresponding author, e-mail address: xuy@fjut.edu.cn%
}, Jieqin Lan, Yin Bai, Weiwei Gao}

\maketitle

\begin{flushleft}
School of Mathematics and Physics, Fujian University of Technology, Fuzhou 350118, China
\end{flushleft}

\begin{abstract}
A direct WIMP (Weakly Interacting Massive Particle) detector with a
neutron veto system is designed to better reject neutrons. An
experimental configuration is studied in the present paper: 984 Ge
modules are placed inside a reactor neutrino detector. In order to discriminate between
nuclear and electron recoil, both ionization and heat signatures are measured using
cryogenic germanium detectors in this detection. The neutrino
detector is used as a neutron veto device.The neutron background
for the experimental design has been estimated using the Geant4
simulation. The results show that the neutron background can
decrease to O(0.01) events per year per tonne of high purity
Germanium.  We calculate the sensitivity to spin-independent WIMP-nucleon elastic
scattering. An exposure of one tonne $\times$ year could reach a
cross-section of about 2$\times$$10^{-11}$ pb.
\end{abstract}


\begin{flushleft}
\end{flushleft}

\section{Introduction}

In direct searches for WIMPs, there are three different methods used
to detect the nuclear recoils, including collecting ionization,
scintillation and heat signatures induced by them. The background of
this detection is made up of electron recoils produced by $\gamma$
and $\beta$ scattering off electrons, and nuclear recoils produced
by neutrons scattering elastically off target nuclei. Nuclear recoils
can be efficiently discriminated from electron recoils with
pulse shape discrimination, hybrid measurements and so on. The
rejection powers of these techniques can even reach  $\>$$10^6$
\cite{CDMSII2010, LAr2008}. For example, the CDMS-II
\cite{CDMSII2010} and EDELWEISS-II \cite{EDELWEISS-II2011}
experiments measure both ionization and heat signatures using
cryogenic germanium detectors in order to discriminate between
nuclear and electron recoils, and the XENON100 \cite{XENON2012} and
ZEPLIN-III \cite{ZEPLIN-III2011} experiments measure both ionization
and scintillation signatures using two-phase xenon detectors.
However, it is very difficult to discriminate between nuclear
recoils induced by WIMPs and by neutrons. This discrimination and
reduction of neutron backgrounds are the most important tasks in direct dark matter searches.
\par
The cross-sections of neutron-nuclei interactions are much larger
than those of WIMP-nuclei, so the multi-interactions between
neutrons and detector components are applied to tag neutrons and
thus separate WIMPs from neutrons. In the ZEPLIN-III experiment, the
$0.5\%$ gadolinium (Gd) doped polypropylene is used as the neutron
veto device, and its maximum tagging efficiency for neutrons reaches
about $80\%$ \cite{ZEPLIN-III2010}. In Ref. \cite{GdWater2008}, the
$2\%$ Gd-doped water is used as the neutron veto, and its neutron
background can be reduced to 2.2 (1) events per year per tonne of
liquid xenon (liquid argon). In our past work \cite{GdLS2010}, the
reactor neutrino detector with $1\%$ Gd-doped liquid scintillator
(Gd-LS) is used as the neutron veto system, and its neutron
background can be reduced to about 0.3 per year per tonne of liquid
xenon. These neutron background events are mainly from the
spontaneous fission and ($\alpha$, n) reactions due to $^{238}U$ and
$^{232}Th$ in the photomultiplier tubes (PMTs) in the liquid xenon.
\par
Because of its advantages of the low background rate, energy
resolutions and low energy threshold, high purity Germanium (HPGe)
is widely applied in dark matter and neutrino-less double beta decay
experiments\cite{CDMSII2010,EDELWEISS-II2011,Gerda,cdex,CoGeNT}.
In our work, the cryogenic $^{73}Ge$ is used as a WIMP target material
and WIMPs are detected by both ionization and heat channels
(like the CDMS-II and EDELWEISS-II experiments). A detector configuration that
can shield and tag neutrons will better reject neutron background in
dark matter experiments. The feasibility of direct WIMPs detection
with the neutron veto based on the neutrino detector had been
validated in our past work\cite{GdLS2010}. So, in the present paper,
a neutrino detector with Gd-LS ($1\%$ Gd-doped) is still used as a
neutron-tagged device and WIMP detectors with HPGe targets(called Ge
modules) are placed inside the Gd-LS. Here we designed an
experimental configuration: 984 Ge modules are individually placed
inside four reactor neutrino detector modules which are used as a
neutron veto system. The experimental hall of the configuration is
assumed to be located in an underground laboratory with a depth of
910 meter water equivalent (m.w.e.), which is similar to the far
hall in the Daya Bay reactor neutrino
experiment\cite{DayabayProp2007}.
The neutron background for this design are estimated using the
Geant4\cite{geant4} simulation.
\par
The basic detector layout will be described in the section 2. Some
features of the simulation in our work will be described in the
section 3. The neutron background of the experimental configuration
will be estimated in the section 4. The other background will
be roughly estimated in the section 5. We give a conclusion
in the section 6.

\section{Detector description}
\begin{figure}
 \centering
 \includegraphics[width=0.7\textwidth]{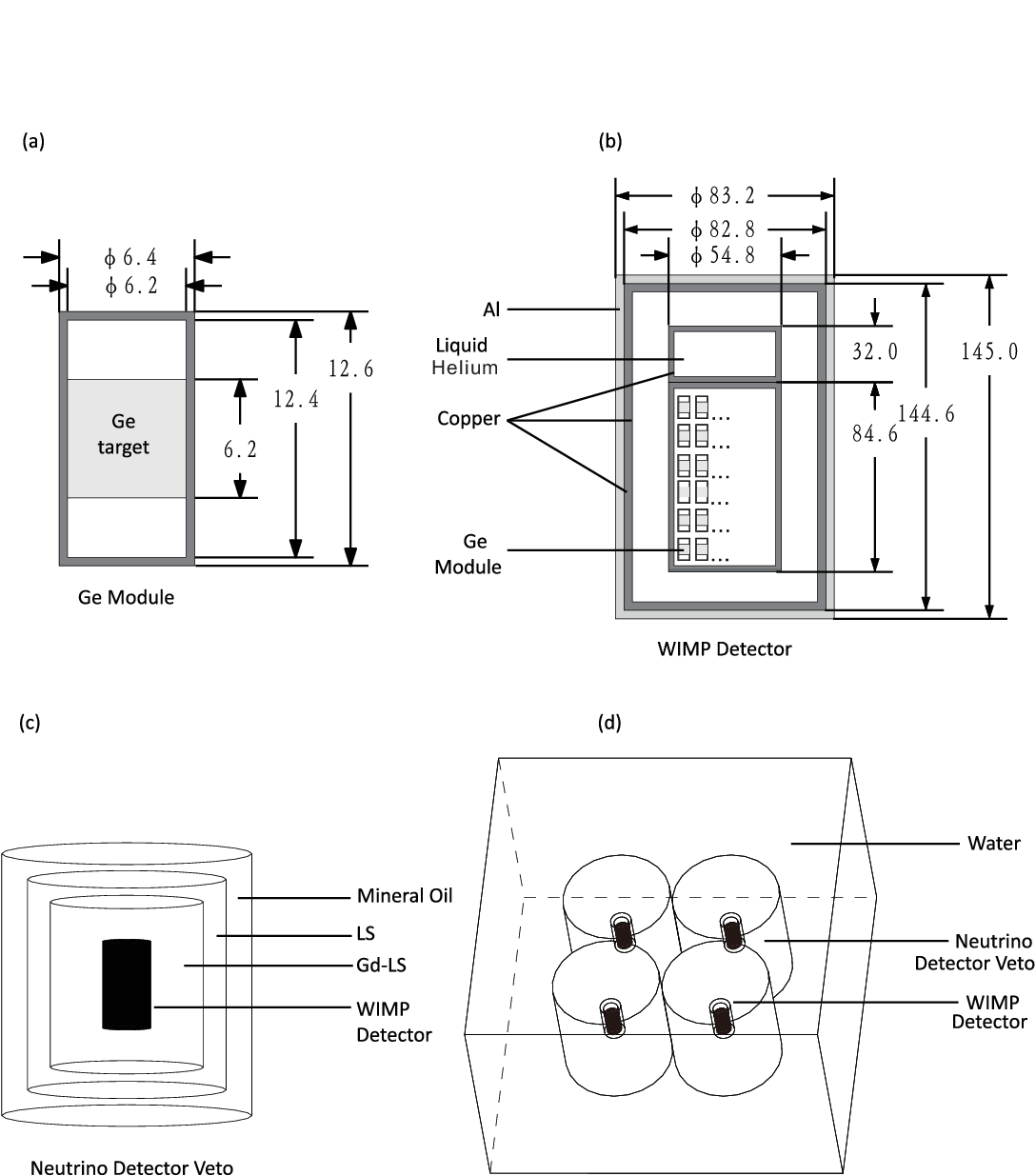}
 \caption{(a): Ge Module with HPGe material(length unit, cm), (b): WIMP
detector with 246 Ge Modules(length unit, cm), (c): a neutrino
detector where a WIMP detector is placed, (d): four WIMP detectors
individually placed inside four neutrino detectors in a water
shield.}
\label{fig:detector}
\end{figure}

Four identical WIMP detectors with HPGe targets are individually
placed inside four identical neutrino detector modules. The
experimental hall of this experimental configuration is assumed to
be located in an underground laboratory with a depth of 910 m.w.e.,
which is similar to the far hall in the Daya Bay reactor neutrino
experiment. The detector is located in a cavern of
20$\times$20$\times$20 $m^3$. The four identical cylindrical
neutrino modules (each 413.6 cm high and 393.6 cm in diameter) are
immersed into a 13$\times$13$\times$8 $m^3$ water pool at a depth of
2.5 meters from the top of the pool and at a distance of 2.5 meters
from each vertical surface of the pool. The detector configuration
is shown in Fig.\ref{fig:detector}.
\par
Each neutrino module is partitioned into three enclosed zones. The
innermost zone is filled with the 1\% Gd doped liquid
scintillator\cite{GdLS2010}(2.6 m height, 2.4 m in diameter), which
is surrounded by a zone filled with unloaded liquid scintillator
(LS)(35cm thickness). The outermost zone is filled with transparent
mineral oil(40cm thickness)\cite{Bayesian2008}. 366 8-inch PMTs are
mounted in the mineral oil. These PMTs are arranged in 8 rings of 30
PMTs on the lateral surface of the oil region, and 5 rings of 24,
18, 12, 6, 3 on the top and bottom caps.
\par
Each WIMP detector consists of a outer copper vessel(144.6 cm
height, 82.8 cm in diameter and 0.8 cm thickness) which is
surrounded a Aluminum(Al) reflector(0.2cm thickness) and inner
copper vessel(116.6 cm height, 54.8 cm in diameter and 0.5 cm
thickness). There is a vacuum zone between the outer and inner
copper vessels (about 13 cm thickness). The part inside the inner
copper vessel is made up of two components: the upper component is a
cooling system with liquid Helium of very high purity(32 cm
height, 54.8 cm in diameter) and the lower one is an active target
of 246 Ge modules arranged in 6 columns (each column includes 4
rings of 20, 14, 6, 1). Each Ge module is made up of a copper vessel
and a HPGe target: there is a HPGe target(6.2 cm height, 6.2 cm in
diameter, $\sim$1 kg) in a 0.1 cm thick copper vessel (12.6 cm
height, 6.4 cm in diameter).
\section{Some features of simulation}
\par
The Geant4 (version 8.2) package\cite{geant4} has been used in our
simulations. The physics list in the simulations includes
transportation processes, decay processes, low energy processes,
electromagnetic interactions (multiple scattering processes,
ionization processes, scintillation processes, optical processes,
cherenkov processes, Bremsstrahlung processes, etc.) and hadronic
interactions (lepton nuclear processes, fission processes, elastic
scattering processes, inelastic scattering processes, capture
processes, etc.). The cuts for the productions of gammas, electrons
and positrons are 1 mm, 100 $\mu$m and 100$\mu$m, respectively. The
quenching factor is defined as the ratio of the detector response to
nuclear and electron recoils. The Birks factor for protons in the
Gd-LS is set to 0.01 g/cm$^{2}$/MeV, corresponding to the quenching
factor 0.17 at 1 MeV, in our simulation.
\section{Neutron background estimation}
In order to reject neutrino
background, the recoil energy was set to a range from 10 keV to 100
keV in this work. Multi-scattering with the detector, neutrons can
be tagged by the Ge modules each other. But there is deposit energy
in only a Ge module for a WIMP interaction. Proton recoils induced
by neutrons and neutron-captured signals are used to tag neutrons
which reach the Gd-LS. The energy deposition produced by proton
recoils is close to a uniform distribution. Neutrons captured on Gd
and H lead to a release of about 8 MeV and 2.2 MeV of $\gamma$
particles, respectively. Due to the instrumental limitations of the
Gd-LS, we assume neutrons will be tagged if their energy deposition
in the Gd-LS is more than 1 MeV, corresponding to 0.17 MeVee
(electron equivalent energy). In the Gd-LS, it is difficult to
distinguish signals induced by neutrons from electron recoils, which
are caused by the radioactivities in the detector components and the
surrounding rocks. But these radioactivities can be controlled to
less than $\sim$ 50 Hz according to the Daya Bay
experiment\cite{DayabayProp2007}. If we assume a 100 $\mu$s for
neutron tagging time window, the indistinguishable signals due to
the radioactivities will result in a total dead time of less than 44
hours per year.
\par
Neutrons are produced from the detector components and their
surrounding rock. For the neutrons from the surrounding rock there
are two origins: first by spontaneous fission and ($\alpha$, n)
reactions due to $\rm{^{238}U}$ and $\rm{^{232}Th}$ in the rock (these neutrons can be omitted
because they are efficiently shielded, see Sec.4.2), and secondly by
cosmic muon interactions with the surrounding rock.
\par
We estimated the number of neutron background in the Ge target of
one tonne. This number has been normalized to one year of data
taking and are summarized in Tab.\ref{tab:bg}.
\subsection{Neutron background from detector components}
Neutrons from the detector components are induced by ($\alpha$, n)
 reactions due to U and Th. According to Mei et al.\cite{Mei2009}, the
 differential spectra of neutron yield can be expressed as
\begin{center}
 $\displaystyle Y_{i}(E_{n})=N_i{
\sum_{j}\frac{R_{\alpha}(E_{j})}{S_{i}^{m}(E_{j})}}$$\displaystyle
\intop_{0}^{E_{j}}\frac{d\sigma(E_{\alpha},E_{n})}{dE_{\alpha}}dE_{\alpha}$
\end{center}
where $N_i$ is the total number of atoms for the $i^{th}$ element in
the host material, $R_\alpha$$(E_j)$ refers to the $\alpha$-particle
production rate for the decay with the energy $E_j$ from $^{232}Th$
or $^{238}U$ decay chain, $E_{\alpha}$ refers to the $\alpha$
energy, $E_n$ refers to the neutron energy, and $S_{i}^{m}$ is the
mass stopping power of the $i^{th}$ element.
\subsubsection{Neutrons from copper vessels}
In the copper vessels, neutrons are produced by the U and Th
contaminations and emitted with their average energy of 0.81
MeV\cite{Mei2009}. Their total volume is about $5.4\times10^4$
$cm^3$. The radioactive impurities Th can be reduced to
$2.5\times10^{-4}$ ppb in some copper samples\cite{Martin2009}
 If we conservatively assume a 0.001 ppb U/Th
concentrations in the copper material\cite{Exo200_2007}, a rate of
one neutron emitted per $4\times10^4$ $cm^3$ per year is
estimated\cite{GdWater2008}. Consequently, there are 1.3 neutrons
produced by the all copper vessels per year.
\par
The simulation result is summarized in Tab.\ref{tab:bg}. 0.39
neutron events/(ton$\cdot$yr) reach the HPGe targets, their energy
deposition falls in the same range as that of the WIMP interactions
and there is deposit energy in only a Ge module (see
Tab.\ref{tab:bg}). As 0.01 of them are not tagged in the Gd-LS,
these background events cannot be eliminated. The uncertainty of the
neutron background from the copper vessels are from the binned
neutron spectra in the Ref.\cite{Mei2009}. But the neutron
background errors from the statistical fluctuation (their relative
errors are less than 1$\%$) are too small to be taken into account.
\subsubsection{Neutrons from front-end electronics}
The U and Th contaminations in the Copper material are considered as
the only neutron source in the front-end electronics in the Ge
modules. If we assume that a 2 ppb U/Th concentrations in the copper
material and their total volume with about 500 $cm^3$, there are 25
neutrons produced by the all front-end electronics per year.
\par
The simulation result is summarized in Tab.\ref{tab:bg}. 1.70
neutron events/(ton$\cdot$yr) reach the HPGe targets, their energy
deposition falls in the same range as that of the WIMP interactions
and there is deposit energy in only a Ge module (see
Tab.\ref{tab:bg}). As 0.038 of them are not tagged in the Gd-LS,
these background events cannot be eliminated. The uncertainty of the
neutron background from the copper vessels are from the binned
neutron spectra in the Ref.\cite{Mei2009}. But the neutron
background errors from the statistical fluctuation (their relative
errors are less than 1$\%$) are too small to be taken into account.
\subsubsection{Neutrons from other components}
The U and Th contaminations in other detector components also
contribute to the neutron background in our experiment setup.
Neutrons from the aluminum reflectors are emitted with the average
energy of 1.96 MeV\cite{Mei2009}. The U and Th contaminations in the
carbon material are considered as the only neutron source in the
Gd-LS/LS. Neutrons from the Gd-LS/LS are emitted with the average
energy of 5.23 MeV\cite{Mei2009}. The U and Th contaminations in the
quartz material are considered as the only neutron source in the
PMTs in the oil. Neutrons from PMTs are emitted with the average
energy of 2.68 MeV\cite{Mei2009}. The U and Th contaminations in the
iron material are considered as the only neutron source in the
stainless steel tanks. Neutrons from the stainless steel tanks are
emitted with the average energy of 1.55 MeV\cite{Mei2009}. We
evaluated the neutron background from the above components using the
Geant4 simulation. All the nuclear recoils in the HPGe targets,
which is in the same range as that of the WIMP interactions and and
there is deposit energy in only a Ge module, are tagged. The neutron
background from these components can be ignored.
\subsection{Neutron background from natural radioactivity in the surrounding rock}
In the surrounding rock, almost all the neutrons due to natural
radioactivity are below 10 MeV \cite{GdWater2008, Carson2004}. Water
can be used for shielding neutrons effectively, especially in the
low energy range of less than 10 MeV \cite{Carmona2004}. The Ge
detectors are surrounded by about 2.5 meters of water and more than
1 meter of Gd-LS/LS, so these shields can reduce the neutron
contamination from the radioactivities to a negligible level.
\subsection{Neutron background due to cosmic muons}
Neutrons produced by cosmic muon interactions constitute an
important background component for dark matter searches. These
neutrons with a hard energy spectrum extending to several GeV
energies, are able to travel far from produced vertices.
\par
\begin{figure}
 \centering
 \includegraphics[width=0.8\textwidth]{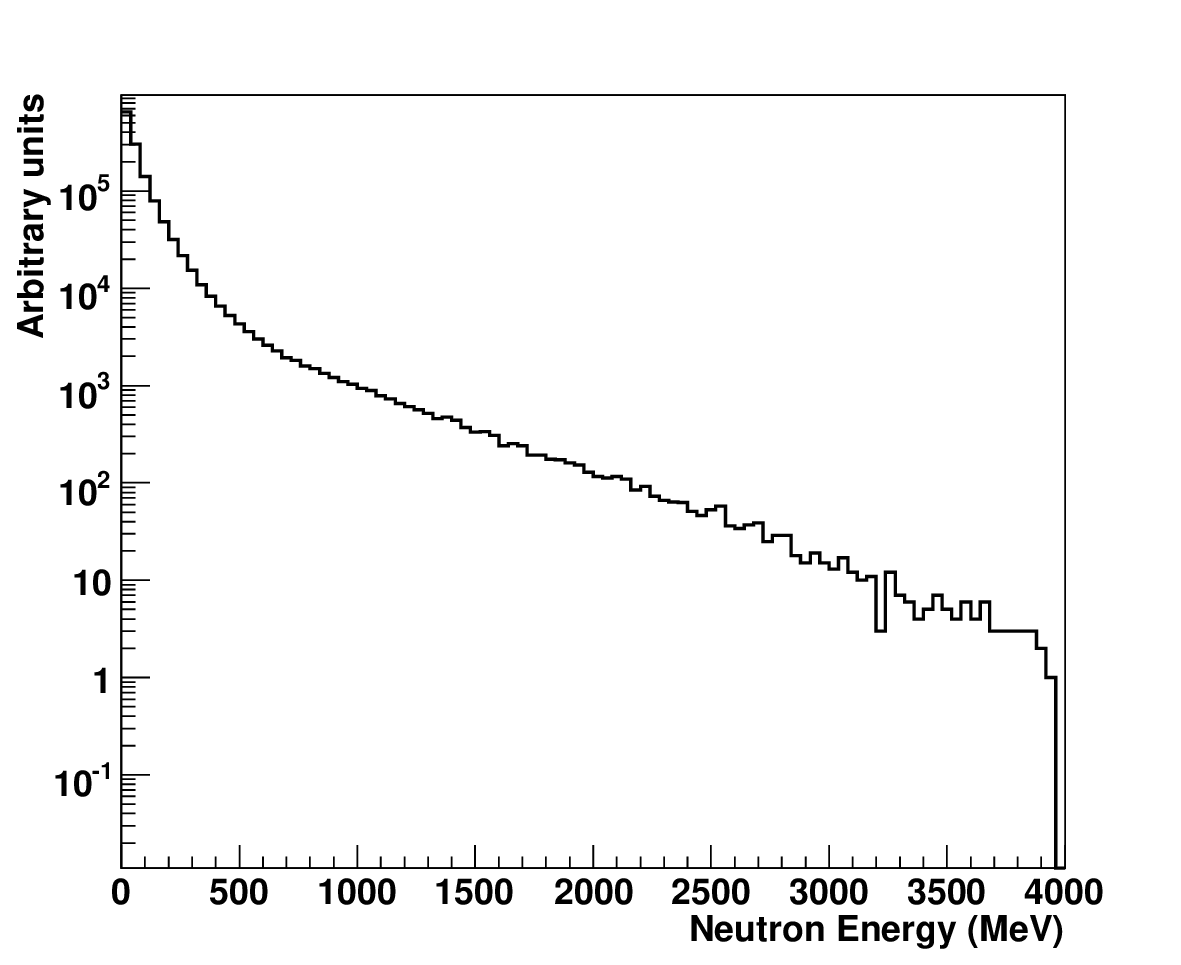}
 \caption{The energy spectrum of cosmogenic neutrons at depth of 910 m.w.e.}
 \label{fig:energy}
\end{figure}

The total cosmogenic neutron flux at a depth of 910 m.w.e. is
evaluated by a function of the depth for a site with a flat rock
overburden \cite{Mei2006}, and it is 1.31$\times$$10^{-7}$
$cm^{-2}s^{-1}$ . The energy spectrum (see Fig.\ref{fig:energy}) and
angular distribution of these neutrons are evaluated at the depth of
910 m.w.e. by the method in \cite{Mei2006, Wang2001}. The neutrons
with the specified energy and angular distributions are sampled on
the surface of the cavern, and the neutron interactions with the
detector are simulated with the Geant4 package. Tab.\ref{tab:bg}
shows that 31.5 neutron events/(ton$\cdot$yr) reach the HPGe targets,
their energy deposition is in the same range as that of the WIMP
interactions and there is deposit energy in only a Ge module. 0.24 of
them are not tagged by the Gd-LS/LS. Muon veto systems can tag muons
very effectively, thereby most cosmogenic neutrons can be rejected.
The water Cherenkov detector in our simulation are used to tag
cosmic muons and then reject them. These detectors are similar to
the ones in the Daya Bay experiment, and the muon rejection is
consistent with the result of the Daya Bay experiment, that is the
contamination level can even be reduced by a factor of about
30\cite{DayabayProp2007}. This could lead to the decrease of
cosmogenic neutron contamination to 0.008 events/(ton$\cdot$yr). The
uncertainties of the cosmogenic neutron background in
Tab.\ref{tab:bg} are from the statistical fluctuation.

\section{Rough estimation of other background}
Besides neutron background, other background events are mainly from reactor
neutrino events and electron recoils in the experimental design in
the present paper. The contamination caused by electron recoils
consists of bulk electron recoil events and surface events.
\subsection{Contamination due to reactor neutrino events}
Since neutrino detectors are fairly close to nuclear reactors (about
2 kilometers away) in reactor neutrino experiments, a large number
of reactor neutrinos will pass through the detectors, and nuclear
recoils will be produced by neutrino elastic scattering off target
nucleus in the WIMP detectors. Although neutrinos may be a source of
background for dark matter searches, they can be reduced to a
negligible level by setting the recoil energy threshold of 10
keV\cite{Jocely2007}. Besides, nuclear recoils may also be produced
by low energy neutrons produced by the inverse $\beta$-decay
reaction $\bar{\nu_{e}}+p\rightarrow e^{+}+n$. But their kinetic
energies are almost below 100 keV\cite{chooz2003}, and their maximum
energy deposition in the WIMP detectors is as large as a few keV.
Thus the neutron contamination can be reduced to a negligible level
by the energy threshold of 10 keV.
\subsection{Contamination due to electron recoils}
Nuclear recoils can be efficiently discriminated from electron recoils with hybrid measurements.
The cryogenic Ge detectors measure both ionization and heat signatures,
and the rejection power against electron recoil events can reach  $\>$$10^6$\cite{CDMSII2010}.
The surviving electron recoils are mainly from surface events by this technique.
Particle interactions may suffer from a suppressed ionization signal if the interactions occur
in the first few microns of the crystal surfaces. For events interacting in the first few microns
the ionization loss is sufficient to misidentified these as nuclear recoils. And they are referred to
as surface events, which mainly occur due to radioactive contamination on Ge crystal surfaces\cite{cdmsii2008}.
The radioisotope $\rm{^{210}Pb}$ is a major component of the surface events,
which plate-out from radon exposure during detector production\cite{spcdms2006}.  Here we assume that the
surface event rate is about 1 counts/Ge module/day, according to the ref.\cite{cdmsiitosuper2010}.
Considering the rejection power against electron recoils with O($10^6$), we
roughly estimate that the surface event contamination is about 0.4 events/(ton$\cdot$yr).
\section{Conclusion and discussion}
The neutron background can be effectively suppressed by the neutrino
detector used as the neutron veto system in direct dark matter
searches. Tab.\ref{tab:bg} shows the total neutron contamination are
0.056 events/(ton$\cdot$yr). And compared to Ref.\cite{GdLS2010}, it
is reduced by a factor of about 5. This decrease is caused by the
reason that the neutron contamination is mainly from the PMTs in the
Xenon detector, but there are no photomultiplier tubes (PMTs) in the
HPGe detector. According to our work, the neutron background is
mainly from its front-end electronics in this configuration with the
HPGe targets. Compared to electron recoils\cite{CoGeNT}, the
estimated neutron contamination in the paper can be ignored. After
finishing a precision measurement of the neutrino mixing angle
$\theta_{13}$, we can utilize the existing experiment hall and
neutrino detectors. This will not only save substantial cost and
time for direct dark matter searches, but the neutron background
could also decrease to O(0.01) events per year per tonne of HPGe in
the case of the Daya Bay experiment. According to
Ref.\cite{Mei2006}, The neutron fluxes in the RENO (in an
underground laboratory with a depth of 450 m.w.e.), Double CHOOZ (in
an underground laboratory with a depth of 300 m.w.e.)
experiments\cite{reno,dc} are respectively about 5 and 3 times more than that of
the Daya Bay experiment. So their neutron backgrounds are roughly
estimated to be about 0.1 events/(ton$\cdot$yr), if their detector
configurations are the same as the one described above. In the case
of the CDMSII (in an underground laboratory with a depth of 2100
m.w.e.)\cite{CDMSII2010}, its neutron flux is reduced by a factor of
about 10. If its detector is the same as the one described above,
its neutron background is roughly estimated to be about 0.05
events/(ton$\cdot$yr) (its cosmogenic neutron background can be
ignored.)
\par
To evaluate the detector capability of directly detecting dark
matter, we assume a standard dark matter galactic
halo\cite{Lewin1996}, an energy resolution that amounts to 25$\%$
for the energy range of interest and 32$\%$ nuclear recoil
acceptance\cite{cdms2009}.
\par
\begin{figure}
 \centering
 \includegraphics[width=0.8\textwidth]{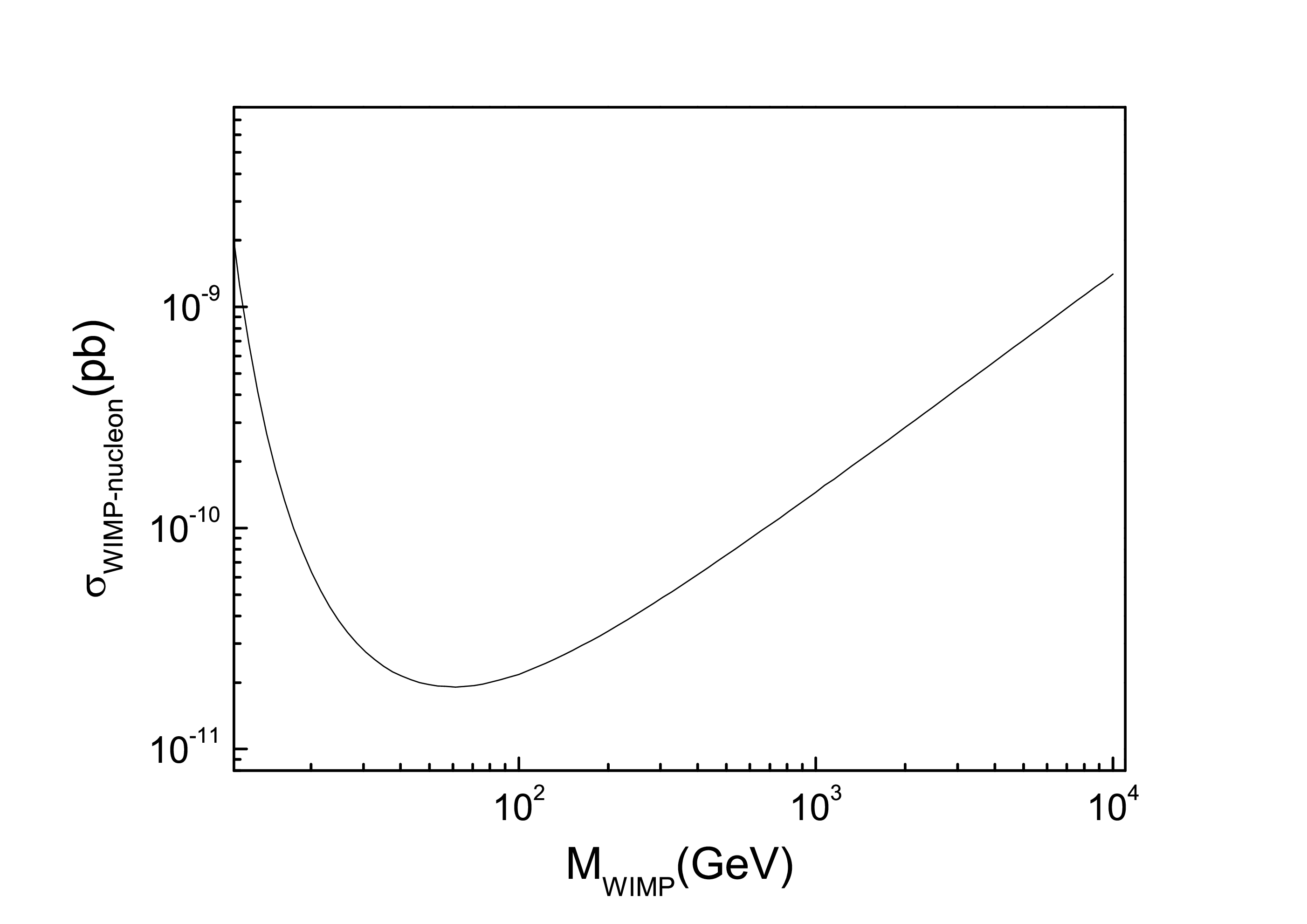}
 \caption{We calculate the sensitivity to spin-independent
WIMP-nucleon elastic scattering assuming an exposure of one tonne
$\times$ year. The calculation shows this exposure could reach a
cross-section of about 2$\times$$10^{-11}pb$ at the 90$\%$
confidence level. The tool from Ref.\cite{limits} has been used. }
 \label{fig:limit}
\end{figure}

If no signals are significantly observed, a sensitivity to WIMP-nucleon
spin-independent elastic scattering can be calculated via the same
method as Ref.\cite{GR1998}. Our calculation shows that an exposure
of one tonne $\times$ year could reach a cross-section of about
2$\times$$10^{-11}pb$ at the 90$\%$ confidence level (see
Fig.\ref{fig:limit}).
\section{Acknowledgements}
This work was supported by the National Natural Science Foundation
of China (NSFC) under the contract No. 11235006, the Science Fund of
Fujian University of Technology under the contract No. GY-Z14061 and the Natural Science Foundation of
Fujian Province in China under the contract No. 2015J01577.
\par

\newpage
\begin{table}

  \centering
  \begin{tabular}{|c|c|c|}
  \hline
  & 10keV<$E_{recoil}$<100keV and only a Ge module & Not Tagged\\
  \hline
  copper vessels & 0.39$\pm$0.02 & 0.01$\pm$0.004\\
  \hline
  front-end electronics & 1.70$\pm$0.06 & 0.038$\pm$0.01\\
  \hline
  cosmic muons & 31.5$\pm$1.59  & 0.24$\pm$0.14\\
  \hline
  muon veto & 1.05$\pm$0.29 & 0.008$\pm$0.025\\
  \hline
  total(muon veto) & 3.1$\pm$0.30 & 0.056$\pm$0.027\\
  \hline
  \end{tabular}
  \caption{Estimation of neutron background from different sources for
an underground laboratory at a depth of 910 m.w.e\@. The column
labeled "10keV<$E_{recoil}$<100keV and only a Ge module" identifies
the number of neutrons whose energy deposition in the Ge is in the
same range as WIMP interactions and there is deposit energy in only
a Ge module. The column labeled "Not Tagged" identifies the number
of neutrons which are misidentified as WIMP signatures (their energy
deposition in the Ge is in the same range as WIMP interactions while
their recoil energies in the Gd-LS/LS are less than the energy
threshold of 1 MeV). The row labeled "copper vessel" identifies the
number of neutrons from the copper vessels. The row labeled
"front-end electronics" identifies the number of neutrons from the
front-end electronics. The row labeled "cosmic muons" identifies the
number of cosmogenic neutrons in the case of not using the muon veto
system. The row labeled "muon veto" identifies the number of
cosmogenic neutrons in the case of using the muon veto system. We
assume that neutron contamination level from cosmic muons decreases
by a factor of 30 using a muon veto system. Only the total
background in the case of using the muon veto system is listed in
this table. The terms after $\pm$ are errors.}
\label{tab:bg}
\end{table}
\newpage
\newpage
\newpage


\newpage
\begin{thebibliography}{}
\bibitem{CDMSII2010} Z. Ahmed, et al., Science 327, 1619 (2010)
\bibitem{LAr2008} W.H.Lippincott, et al., Phys. Rev. C78, 035801
(2008)
\bibitem{EDELWEISS-II2011} E. Armengaud, et al., Phys. Lett. B702, 329
(2011)
\bibitem{XENON2012} E. Aprile, et al., Astropart. Phys. 35, 573
(2012)
\bibitem{ZEPLIN-III2011} D. Akimov, et al., Phys. Lett. B709, 14 (2012)
\bibitem{ZEPLIN-III2010} D. Akimov, et al., Astropart, Phys. 34, 151 (2010)
\bibitem{GdWater2008} A. Bueno, M. C. Carmona and A. J. Melgarejo, JCAP. 08, 019
(2008).
\bibitem{GdLS2010} Ye Xu, et al., JCAP06, 009 (2011)
\bibitem{Gerda} I. Abt, M.Altmann, et al., Nucl. Phys. B Proceedings Supplements 145, 242-245 (2005)
\bibitem{cdex} Qian Yue and Henry T. Wong, Dark Matter Search with
sub-keV Germanium Detectors at the China Jinping Underground
Laboratory, arXiv: 1201.5373
\bibitem{CoGeNT} C.E.Aalseth, et al., Phys. Rev. Letts. 107, 141301 (2011)
\bibitem{DayabayProp2007} Daya Bay Collaboration,
arXiv:hep-ex/0701029v1; F.P. An, et al., Phys. Rev. Letts. 108,
171803 (2012)
\bibitem{Bayesian2008} Ye Xu, et al., Nucl. Meth. and Instr. A 592,
451-455 (2008)
\bibitem{geant4} S. Agostinelli, et al., Nucl. Instru. Meth. A506, 250
(2003)
\bibitem{Mei2009} D. M. Mei, C. Zhang and A. Hime, Nucl.
Instrum. Meth. A606, 651 (2009)
\bibitem{Martin2009} Marin E. Keillor, et al., J. Radioanal. Nucl. Chem. 282, 703-708 (2009)
\bibitem{Exo200_2007} D. S. Leonard, et al., Nucl. Instrum. Meth.
A591, 490 (2008)
\bibitem{Carson2004} M. J. Carson, et al., Astropart. Phys. 21, 667 (2004)
\bibitem{Carmona2004} J. M. Carmona, et al., Astropart. Phys. 21, 523 (2004)
\bibitem{Mei2006} D. M. Mei and A. Hime, Phys. Rev. D73, 053004 (2006)
\bibitem{Wang2001} Y.F.Wang, et al., Phys. Rev.D64, 013012 (2001)
\bibitem{Jocely2007} Jocelyn Monroe and Peter Fisher, Phys. Rev. D76, 033007
(2007)
\bibitem{chooz2003}M. Apollonio, et al., CHOOZ Collaboration, Eur. Phys.
J. C27, 331-374(2003), arXiv: hep-ex/0301017
\bibitem{cdmsii2008}T. Bruch, for the CDMS Collaboration, the Proc. of the 4th PATRAS Workshop on Axions, WIMPs and WISPs DESY, 18-21 June 2008, arXiv:0809.4186
\bibitem{spcdms2006}D.S.Akerib, et al., the CDMS Collaboration, NIMA 559,411-413 (2006)
\bibitem{cdmsiitosuper2010}T. Bruch, for the CDMS Collaboration, the Proc. of the 5th Patras Workshop on Axions, WIMPs and WISPs, Durham, England, 13-17 July 2009, arXiv:1001.3037
\bibitem{reno}J.K. Ahn, et al., Phys. Rev. Letts. 108, 191802 (2012)
\bibitem{dc}F. Ardellier, et al., Double Chooz: A Search for the Neutrino Mixing Angle
$\theta_{13}$, arXiv: hep-ex/0606025
\bibitem{Lewin1996} J. D. Lewin and P. F. Smith, Astropart. Phys. 6, 87 (1996)
\bibitem{cdms2009}Z. Ahmed, et al., the CDMS Collaboration, Science 327, 1619-1621 (2010), arXiv:0912.3592
\bibitem{GR1998} G.J.Feldman and R.D.Cousins, Phys. Rev. D57,
3873 (1998)
\bibitem{limits}http://pisrv0.pit.physik.uni-tuebingen.de/darkmatter/limits/index.php
\end{thebibliography}
\end{document}